\documentstyle[12pt,epsf]{article}

\newcommand{\re}[1]{(\ref{#1})}

\newcommand{\bfr}{\begin{flushright}}
\newcommand{\bfl}{\begin{flushleft}}
\newcommand{\efl}{\end{flushleft}}
\newcommand{\efr}{\end{flushright}}
\newcommand{\bc}{\begin{center}}
\newcommand{\ec}{\end{center}}
\newcommand{\be}{\begin{equation}}
\newcommand{\ee}{\end{equation}}
\newcommand{\bea}{\begin{eqnarray}}
\newcommand{\eea}{\end{eqnarray}}
\newcommand{\ba}{\begin{array}}
\newcommand{\ea}{\end{array}}
\newcommand{\edc}{\end{document}}

\newcommand{\ds}{\displaystyle}
\newcommand{\dsf}{\displaystyle\frac}

\newcommand{\il}{\int\limits}
\newcommand{\pal}{\partial}
\newcommand{\pirow}{\pi\rho\omega}
\newcommand{\pirows}{\pi\rho\omega\sigma}
\newcommand{\piros}{\pi\rho\omega\sigma}
\newcommand{\msig}{m_{\sigma}}
\newcommand{\fpi}{f_{\pi}}
\newcommand{\cg}{C_{g}}
\newcommand{\vecr}{\vec{r}}

\newcommand{\lab}{\label}
\newcommand{\ci}{\cite}
\newcommand{\Tr}{\mbox{Tr}}
\newcommand{\mpi}{m_{\pi}}
\newcommand{\ga}{g_{A}}
\newcommand{\vnn}{V_{NN}}
\newcommand{\vs}{\vspace{-0.25cm}}
\newcommand{\sigmat}{\tilde{\sigma}}

\begin{document}
\begin{center}

\hfill{\bf rerevised version}

\vspace{2 cm}

\begin{center}

{{\Large\bf
The nucleon--nucleon interaction and properties of
the\\[0.1em] nucleon in a $\pirow$  soliton model including a dilaton
field\\[0.3em] with anomalous dimension}}

\vspace{2 cm}

Ulf-G. Mei{\ss}ner
\footnote{E-mail: Ulf-G.Meissner@fz-juelich.de},
A. Rakhimov
\footnote{E-mail: abdulla@iaph.silk.org; Permanent address:
Institute of Nuclear Physics, Academy of Sciences,
    Usbekistan (CIS)},
U. Yakhshiev$^a$\footnote{E-mail: ulug@iaph.silk.org;
Permanent address: Institute of
Applied Physics, Tashkent State University,
Tashkent-174, Usbekistan (CIS)  
}\\[1cm]

{\em{
Forschungszentrum J{\" u}lich, Institut f{\" u}r Kernphysik
(Theorie)\\
D-52425  J{\" u}lich, Germany}}

\bigskip

{\em {
$^a$International Centre for Theoretical Physics (ICTP)\\
34014, Trieste, Italy
}}

\end{center}
\thispagestyle{empty}

\vspace{2cm}

\begin{abstract}
\indent

 We investigate an extended chiral soliton model which includes
$\pi, \rho, \omega $ and $\sigma $ mesons as explicit degrees of freedom.
The Lagrangian incorporates chiral symmetry and broken scale invariance.
A scalar--isoscalar meson $\sigma$ is associated with a quarkonium
dilaton field with a mass $\msig\approx 550\,$MeV.
We show that the scalar field with anomalous dimension
slightly changes the static and electromagnetic properties of the nucleon.
In contrast, it plays a significant role
in  nucleon--nucleon dynamics and gives an opportunity
to describe well the two--nucleon interaction.
\end{abstract}
\vspace {1cm}
Keywords: [
Skyrme model, sigma meson, scale dimension,
nucleon properties, meson-nucleon form factors,
 nucleon-nucleon interaction.]
\end{center}


\newpage

\addtocounter{page}{-1}
\section{Introduction}
\indent

In ref.\cite{fts} Furnstahl, Tang and Serot (FTS) proposed a
new model for nuclear matter
and finite nuclei that realizes QCD symmetries such as chiral symmetry,
broken scale invariance and the phenomenology of vector meson dominance.
An important feature of this approach is the inclusion of light scalar
 degrees of freedom,  which are given an anomalous scale dimension.
The vacuum dynamics of QCD is  constrained by the trace anomaly
and related low--energy theorems of QCD. The scalar--isoscalar sector
of the theory  is divided into a low mass part that is adequately described
by a scalar meson (quarkonium) with anomalous dimension
and a high mass part (gluonium), that can be ``integrated out'',
leading to various couplings among the remaining fields. The application of the
model
to the properties of nuclear matter as well as finite nuclei
gave a satisfactory description.
Further developments of the model~\cite{rhofts,librown} showed
that the light scalar related to  the trace anomaly  can play a significant
role
not only in the description of bound nucleons but also in  the description
of heavy--ion collisions.It was also shown that the anomalous
cannot be due to an effect of nuclear density on the trace anomaly
of QCD \ci{rhofts}.

Here a natural question arises: What is the role of this light quarkonium
in the description of the
properties of a single nucleon,  when it is taken into account
in topological nonlinear chiral soliton models,  which are similar to the FTS
effective Lagrangian on the  single nucleon level?
In the present paper we introduce a dilaton field with
an anomalous dimension into the $\pirow$--model~\cite{meisnpa}
and investigate some properties of single nucleons which emerge
as  solitons in the sector with baryon number one $(B=1)$.

It is well known that a scalar--isoscalar  meson,\footnote{Here
and in what follows, we call it sigma meson for simplicity. Although we assume
it to be some kind of quarkonium state, its precise dynamical nature
is of no direct relevance for the following arguments.} the sigma,
plays an important role in the nucleon--nucleon (NN) interaction especially
within one--boson--exchange (OBE) models~\cite{machrep}.
We remark that the missing medium range attraction was
a long standing puzzle in Skyrme like models. Lately it has been shown
that~\cite{vnnsk} explicit inclusion of a scalar meson into
the Skyrme model produces in a natural way the desired attraction.
But these studies have two shortcomings. Firstly,
the scalar meson used in such models~\cite{vnnsk} has nothing to do
with OBE phenomenology,  as it has a large mass and is identified with
a gluonium state. Secondly,
the Lagrangians used in refs.~\cite{vnnsk}   do not include explicit
omega mesons at all,  which may ``spoil'' the mechanism of the attraction due
to its strong repulsion.\footnote{Of course, the stabilization of the
soliton in refs.\cite{vnnsk} via higher derivative terms also leads to
repulsion in the central NN interaction, but this is less easily be
interpreted as single meson exchange.}
Therefore, it would be quite interesting to investigate the central part of
the NN interaction when the light scalar ($\sigma$) and $\omega$--mesons
are both taken into account explicitly.
This is what is done here. In particular, it is important that if one
is to properly describe the intermediate range attraction in the central
NN interaction, the successful description of the single nucleon properties
within the $\pi\rho\omega$ model should not be destroyed. We also note that
in an soliton approach with explicit regulated two--pion loop graphs one
is able to get the proper intermediate range attraction. In that case, however,
one does not stay within a simple OBE approach any more (as done here) and
also needs to calculate the modifications of the isovector two--pion exchange
to the $\rho$ and so on. For comparison, we mention that recent developments of
the original Bonn OBE potential performed at J\"ulich also include multi--meson
exchanges leading to a renormalization of various interactions, couplings and
cut--off
parameters~\cite{juel}. The model we investigate is related closely to the OBE
approximation of the NN force.

\section{The $\pirows$  model}
\indent

Including a  $\sigma$--meson by means of the scale invariance
and trace anomaly of QCD into the $\pirow$--model~\cite{meisrep} can be done
in terms of the following chiral Lagrangian of the coupled $\pirows$
system,
\bea
\ba{l}
{\cal L}=\dsf{ S_{0}^{2}e^{-2\sigma/d} }{2 } \pal_{\mu}\sigma
\pal^{\mu}\sigma
-\dsf{f_{\pi}^{2}e^{-2\sigma/d }}
{4}\Tr L_{\mu}L^{\mu}
-\dsf{f_{\pi}^{2}e^{-2\sigma/d} }{2}
\Tr[l_{\mu}+ r_{\mu}+ig\vec\tau\vec\rho_{\mu}+ig\omega_{\mu}]^2+\\
\quad\\
+\dsf{3}{2}g\omega_{\mu}B^{\mu}-
\dsf{1}{4}(\omega_{\mu\nu}\omega^{\mu\nu}+
\vec\rho_{\mu\nu}\vec\rho^{\, \, \mu\nu})
+\dsf{f_{\pi}^2m_{\pi}^{2}e^{-3\sigma/d} }{2}\Tr(U-1)-\\
\quad\\
-\dsf{d^{2}S_{0}^2\msig^{2}}{16}\left[1-e^{-4\sigma/d}\left(\dsf{4\sigma}{
d}+1\right)\right], \\
\label{lagr}
\ea
\eea
where the pion fields are parametrized in terms of
$U=\exp{(i\vec\tau \cdot \vec\pi/\fpi)}$ and
$\xi=\sqrt{U}$, left/right--handed currents are given by
$L_{\mu}=U^+\pal_{\mu}U$, $l_{\mu}=\xi^+\pal_{\mu}\xi$,
  $ r_{\mu}=\xi\pal_{\mu}\xi^+, $
and the pertinent vector meson ($\vec{\rho}, \omega$) field strength tensors
are
$\vec\rho_{\mu\nu}=\pal_{\mu}\vec\rho_{\nu}-\pal_{\nu}\vec\rho_{\mu}+
g[\vec\rho_{\mu}\times\vec\rho_{\nu}]$ and
$\omega_{\mu\nu}=\pal_{\mu}\omega_{\nu}-\pal_{\nu}\omega_{\mu}$.
Furthermore, the topological baryon number current is given by
$B^{\mu}=\varepsilon^{\mu\alpha\beta\gamma}
\Tr L_{\alpha}L_{\beta}L_{\gamma}/(24\pi^2)$.

In  Eq.\re{lagr}, $S_0$ is the vacuum expectation value of scalar field
in free space matter,
$\fpi$ is the pion decay constant $(\fpi=93~{\rm MeV})  $
 and $g=g_{\rho\pi\pi}$
 is determined through the KSFR  relation $g=m/\sqrt{2}\fpi $.
 The model assumes the masses
of $\rho$ and $\omega$ mesons to be equal, $m_\rho=m_\omega=m$.
The mass of the $\sigma$ is related
to the gluon condensate in the usual way~\cite{fts,myransky}
 $\msig=2\sqrt{\cg}/(dS_0)$,
where $d$ is the scale dimension of scalar field ($d>1$).
Being ``mapped'' onto the states of a nucleon, the Lagrangian Eq.\re{lagr}
will be similar to the FTS effective Lagrangian.

Nucleons arise as soliton solutions  from the Lagrangian Eq.\re{lagr}
in the sector with baryon number $B=1$.
To construct them one goes through a two step procedure. First,
one finds the classical soliton which has
neither good spin nor good isospin.
Then an adiabatic rotation of the soliton is performed
and it is quantized collectively.\footnote{We again refer the reader 
to ref.~\cite{meisrep} for details.}
The classical soliton follows from Eq.\re{lagr}
by virtue of a spherical symmetrical ans\"atze for the meson fields:
\be
\ba{l}
U(\vec r)=\exp{(i\vec\tau\hat{r}\Theta(r))},  \, \, \,
\quad \rho_i^a=\varepsilon_{iak}\hat r_k\dsf{G(r)}{gr},
\quad \omega_{\mu}(\vec r)=\omega(r)\delta_{\mu 0},  \, \, \,
\sigma(\vec r)=\sigma(r)\,\,.
\lab{ansatz}
\ea\ee
In what follows we call $\Theta(r) $,  $G(r) $,
 $\omega(r) $,  and $\sigma(r)  $
the pion--, $\rho $--,  $\omega $--,  and $\sigma$--meson
profile functions, respectively.
The pertinent boundary conditions to ensure baryon number one
and finite energy are,
$\Theta(0)=\pi,  G(0)=-2,  \omega^{\prime}(0)=\sigma^{\prime}(0)=0,
\Theta(\infty)= G(\infty)=\omega(\infty)=\sigma(\infty)=0$.
To project out baryonic states of good spin and isospin,  we
perform a time--independent SU(2) rotation
\be
\ba{l}
U(\vecr, t)=A(t)U(\vecr)A^{+}(t), \, \,
 \xi(\vecr, t)=A(t)\xi(\vecr)A^{+}(t)\\
 \sigma(\vecr, t)=\sigma(r), \quad
\omega(\vecr, t)=\dsf{\phi(r)}{r}[\vec K \hat r]\\
\vec\tau\cdot
\vec\rho_{0}(\vecr, t)=\dsf{2}{g}A(t)\vec\tau\cdot(\vec K\xi_1(r)+
\hat r\vec K\cdot\hat r \xi_2(r))A^+(t), \\
\vec\tau\cdot\vec\rho_{i}(\vecr, t)=A(t)\vec \tau
\cdot\vec\rho_{i}(\vecr)A^{+}(t)
\lab{excit}
\ea
\ee
with $2\vec{K}$ the angular frequency of the spinning mode of soliton,
$i\vec\tau\cdot\vec K=A^{+}\dot{A}$.
This leads to the time--dependent Lagrange function
\be
{\cal L}(t)=\int d\vecr {\cal L}=-M_{H}(\Theta, G, \omega, \sigma)
+\Lambda(\Theta, G, \omega, \sigma, \phi, \xi_1, \xi_2)\Tr(\dot{A}\dot{A}^+)~.
\ee
Minimizing the classical mass $M_{H}(\Theta, G, \omega, \sigma)$
 leads to the coupled differential equations
for $\Theta, G, \omega$ and $\sigma $
 subject to the aforementioned boundary conditions.
In the spirit of the large $N_c $--expansion,  one then extremizes the
moment of inertia $\Lambda(\Theta, G, \omega, \sigma, \phi, \xi_1, \xi_2) $
 which gives the coupled differential
equations for $\xi_1, \xi_2  $ and $\phi $
 in the presence of the background profiles
$\Theta,  G,  \omega $ and $\sigma  $.
  The pertinent boundary conditions are
$\phi(0)=\phi(\infty)=0, \, \,$
$\xi_1^{\prime}(0)=\xi_1(\infty)=0,\,\,$
$\xi_2^{\prime}(0)=\xi_2(\infty)=0,\,\,  $
$2\xi_1(0)+\xi_2(0)=2. $
The masses of nucleon $M_N $ and the mass of $\Delta$ ,
 $M_{\Delta} $,
are then given by $M_N=M_H+3/8\Lambda $ and $M_\Delta=M_H+3/15\Lambda$.

The electromagnetic form factors are obtained in the usual way~\cite{meisrep}
are:
\be
\ba{l}
G^{S}_E({\vec q}^{\ 2})=-\dsf{4\pi m^2}{3g}\ds\il_0^{\infty}
j_0(qr)\omega(r)e^{-2\sigma/d}r^2dr, \\
G^{S}_M({\vec q}^{\ 2})=-\dsf{2\pi M_Nm^2}{3g\Lambda}
\il_0^{\infty}\dsf{j_1(qr)}{qr}\phi(r)e^{-2\sigma/d}r^2dr, \\
G^{V}_E({\vec q}^{\ 2})=\dsf{4\pi}{\Lambda}\ds\il_0^{\infty}j_0(qr)
\left\{ \dsf{\fpi^2}{3}[4s_{2}^4+(1+2c)\xi_1+
\xi_2]e^{-2\sigma/d}+\dsf{g\phi\Theta^{\prime}s^2}{8\pi^2r^2}\right\}r^2dr,\\
G^{V}_M({\vec q}^{\ 2})=\dsf{8\pi M_N}{3}\ds
 \il_0^{\infty} \dsf{j_1(qr)}{qr}
\left\{ 2\fpi^2[2s_{2}^4-Gc]e^{-2\sigma/d}+
\dsf{3g}{8\pi^2}\omega\Theta^{\prime}s^2\right\}r^2dr,\\
\lab{elff}
\ea
\ee
where $ s=\sin(\Theta),  c=\cos(\Theta) $  and $s_2=\sin(\Theta/2)$.
The normalization is $G^{S}_{E}(0)=G^{V}_{E}(0) =1/2$.
Similarly,  meson--nucleon vertex form factors
may be calculated~\ci{meisff}. Of course, such strong interaction form
factors are model--dependent quantities. In the soliton approach,
however, they arise naturally as Fourier transfroms of the meson
distribution within the extended nucleon. Stated differently, the
soliton acts as a source of an extended meson cloud, which leads to
a meson--nucleon interaction region of finite extension. In momentum
space, this extension can be interpreted as a corresponding form
factor. We mention that such a picture underlies the inclusion of
strong form factors say in OBE models. Coming back to our approach,
these form factors are most easily evaluated in the Breit--frame:
\be
\ba{l}
G^{\pi}(-{\vec q}^{\ 2})=
\dsf{8\pi M_Nf_{\pi}}{3q}({\vec q}^{\ 2}+m_{\pi}^{2})
\ds\il_0^{\infty}j_1( qr)sin(\Theta)r^2dr=\\
=\dsf{8\pi M_Nf_{\pi}}{3}\ds\il_0^{\infty}
\dsf{j_1( qr)}{ qr}
\left[-2\Theta^{\prime}c-\Theta^{\prime\prime}rc+\Theta^{\prime 2}rs+
\dsf{2s}{r}+rm_{\pi}^{2}s\right]r^2 dr, \\
G^{\rho}_E(-{\vec q}^{\ 2})=\dsf{2\pi}{g\Lambda }\ds \il_0^{\infty}
j_0( qr)
\left[-\xi_1^{\prime\prime}-\dsf{2\xi_1^{\prime}}{r}+
m^2\xi_1 -\dsf{\xi_2^{\prime\prime}}{3}-\dsf{2\xi_2^{\prime}}{3r}
 +\dsf{m^2\xi_{2}^{2}}{3}\right]r^2dr, \\
G^{\rho}_M(-{\vec q}^{\ 2})=-\dsf{8\pi M_N}{3g}
\il_0^{\infty}\dsf{j_1( qr)}{ qr}\left[
-G^{\prime\prime}+2G/r^2+m^2G\right]r^2dr, \\
G^{\omega}_E(-{\vec q}^{\ 2})=\ds {4\pi }
\il_0^{\infty}j_0( qr)\left[
\omega^{\prime\prime}+\dsf{2\omega^{\prime}}{r}-m^2\omega\right]r^2dr, \\
G^{\omega}_M(-{\vec q}^{\ 2})=\dsf{2\pi M_N}{\Lambda}
\il_0^{\infty}\dsf{j_1( qr)}{ qr}\left[
\phi^{\prime\prime}-\dsf{2\phi}{r^2}-m^2\phi\right]r^2dr, \\
\lab{ff}
\ea
\ee
where the ``electric'' and ``magnetic'' vector meson--nucleon form factors
are connected  to the Dirac  $F_{1}(t) $ and Pauli form factors
$F_{2}(t) $ through
the following relations:
 $ G^{i}_{E}(t)=F_{1}^{i}(t)+tF_{2}^{i}(t)/4M_{N}^{2} $,
$G^{i}_{M}(t)=F_{1}^{i}(t)+F_{2}^{i}(t)$, $(i=\rho, \omega)$.

\section{Results and discussions.}
\subsection{Static and electromagnetic properties of the nucleon.}
\indent

Using the formulas given above we have
calculated static and electromagnetic
properties of nucleon. As can be seen from Eq.\re{lagr}, the Lagrangian
has no free parameters in the $\pirow$ sector. So,  in actual calculations
the parameters $\mpi,  m,  \fpi$ are fixed at their emperical values,
$\mpi=138$~MeV, $m=m_{\rho}=m_{\omega}=770$~MeV,  $\fpi=93$~MeV,
$ g=m/\sqrt{2}\fpi=5.85$.  In the $\sigma$--meson sector there are
in general three  free parameters:  $\msig$, $S_0$ and
the  scale dimension $d$. The latter has been well studied for
nuclear matter calculations~\ci{fts,rhofts}. In  particular,  it
was shown that for $d{\ge}2$  the much debated Brown--Rho
(BR) scaling may be recovered. Therefore,  assuming that there is no
dependence of $d$ on the density,  we shall use the best value
$d=2.6$ found in refs.~\ci{fts,rhofts}.
The values for $S_0$ - were  found to be $S_0=90.6\div 95.6$  MeV ~\ci{fts} .
 So we put  $S_0=\fpi=93$  MeV. The mass of the $\sigma$ or
equivalently the gluon condensate
$\cg=m_{\sigma}^{2}d^2S_{0}^{2}/4=m_{\sigma}^{2}d^2f_{\pi}^{2}/4$
is  uncertain.  We thus consider
two cases: $\msig=550$~MeV   and   $\msig=720$~MeV in accordance with
recent $\pi\pi$ phase shift analyses~\ci{ishida} and with OBE values.
We stress again that the precise nature of such a scalar--isoscalar
field is not relevant here, only that it should not be a pure gluonium
state. A summary of static nucleon properties obtained in both  cases, i.e.
with $\msig=550~$MeV   and   $\msig=720$~MeV,  is given in Table
\ref{tab1}.
One immediately observes that the nucleon mass is again overestimated.
This may not be regarded as a deficiency, since it is known that
quantum fluctuations tend to decrease the mass substantially.

To estimate the influence of the $\sigma$--meson
we also show the results given by minimal version of $\pirow$ model.
As can be seen from  Table \ref{tab1}, the inclusion
of a light sigma meson into the basic $\pirow$ model just slightly
changes the nucleon mass and its electromagnetic properties.
This may be explained by the fact that
 the role of intermediate scalar--isoscalar meson in gamma--nucleon
interactions is negligible.
In contrast,  the presence of the sigma--meson leads to an enhancement
of axial coupling constant as it was first observed
in ref.~\ci{anov}.
Although, the physical mechanism of this change is not clear,
it   may be understood as  mainly due to modification of meson profiles.
This is in marked contrast to the inclusion of pion loop effects, which
tend to lower the axial coupling even further~\cite{zwm}.
In the present model $\ga=0.88$ for the $\pirow$ and
$\ga=0.95 $ for the $\pirows$ model, respectively. One may expect
that an appropriate inclusion of $\sigma$ meson
into a more complete version of the basic $\pirow$ model might
give the desired value $\ga=1.26$.

\subsection{Meson--nucleon form factors and NN interaction}
\indent

One of the usual ways to calculate the meson--nucleon
interaction potential within topological soliton models is the
so--called product ansatz~\ci{vnnsk}. Within this approximation,
the two--Skyrmion potential as a function of
the relative angles of orientation between the Skyrmions has a
compact form,  and the extraction of the NN potential by projection onto
asymptotic two--nucleon states is  straightforward. This procedure gives
only three nonvanishing channels: the central,  spin--spin and tensor
potential.
At large and intermediate distances, the latter two compare well  with e.g. the
phenomenological Paris potential~\ci{paris}.  The major inadequacy
found in such type of calculations is the lack of an intermediate range
attraction in central potential. Although many remedies have been
proposed,  this result may not be genuine for  Skyrme like models. In fact,
the product ansatz, which is not a solution to the equations of motion,
can only be considered accurate at large distances,  and the failure
of these calculations to reproduce the central range
attraction may simply be the failure of the product approximation
to provide an adequate approximation to the exact
solution.\footnote{An early study giving credit to this line of reasoning
can be found in ref.\cite{UKM}.}
Although the lack of central attraction may be recovered by the inclusion
of a scalar--isoscalar meson, the
inherent ansatz dependence of the trial configuration remains as a major
shortcoming of product approximation~\ci{jjpzahed}.

On the other hand there is another natural way which was first used
by Holzwarth and Machleidt~\ci{holzmach}.  They proposed
to calculate $\vnn$ within OBE model taking coupling
constants and meson--nucleon form factors from a microscopical
model such as the Cloudy Bag model or the Skyrme model. It was shown that
the Skyrme form factor is a soft pion form factor that is compatible with the
$\pi N$
and $NN$ systems. We shall use this strategy to investigate
the $NN$ potential within present model.

The meson nucleon form factors given by well known procedure,
proposed first by Cohen~\ci{meisff} are given in
Eq.\re{ff}. Although they were derived in a microscopical
and consequent way, these form factors could not be directly used
in standard OBE schemes. The reason is that the OBE schemes
 ~\ci{machrep} in momentum space  use form factors
defined for fields propagating on a flat metric, whereas
the definition of form factors in Eq.\re{ff} involve a nontrivial
metric. Hence, before using the latters in OBE scheme
one should modify the procedure by redifining meson fields.
The modification for pion nucleon form factors
in $\pirow$ model is clearly outlined in refs. ~\ci{holzmach}.
Now, applying this procedure in the lagrangian
in Eq. \re{lagr} we get the following $\pi NN$ form factor:
\be
\ba{l}
G^{\pi}(-{\vec q}^{\ 2})=
\dsf{8\pi M_Nf_{\pi}}{3q}({\vec q}^{\ 2}+m_{\pi}^{2})
\ds\il_0^{\infty}j_1( qr)\sqrt{M_T(r)}\sin(\Theta)r^2dr=\\
=\dsf{8\pi M_Nf_{\pi}}{3}\ds\il_0^{\infty}
\dsf{j_1( qr)}{ qr}
\left[-2F^{\prime}(r)-rF^{\prime\prime}(r)+\dsf{2F(r)}{r}+
rm_{\pi}^{2}F(r)\right]r^2 dr, \\
\lab{ffmet}
\ea
\ee
where
\be
\ba{l}
M_T=\left[1+2\tan^2(\Theta/2)\right] e^{-2\sigma/d} ,\\
F(r)=\sqrt{3+\cos^2(\Theta(r))-4\cos(\Theta(r))} e^{-\sigma/d} .
\lab{metr}
\ea
\ee
The influence of metric factor $M_T$ to the pion--nucleon form factor
is illustrated in Fig. 1. It is seen that, without the inclusion
of   $M_T$  the form factor is softer than in OBE models (the dashed line in Fig. 1),
while its inclusion via Eqs. \re{ffmet}, \re{metr}
gives a  behavior closer to OBE models. In fact, a monopole
approximation at small $q^2=t$ of the normalised form factor
 $G^{\pi}(t)/G^{\pi}(0)
\approx\Lambda_{\pi}^{2}/(\Lambda_{\pi}^{2}-t)$
gives $\Lambda_{\pi}=860 $ MeV and  $\Lambda_{\pi}=1100 $ MeV
for $M_T=1$ and $M_T\not=1$ respectively, compared
to its emperical fit : $\Lambda_{\pi}^{OBE}=1300\,$MeV (dotted
line in Fig.1). Note, however, that our results for $\Lambda_{\pi}$
are in line with recent coupled--channel calculations
of the J\"ulich group~\cite{RB}. There, a monopole form factor
with $\Lambda_{\pi} \simeq 800\,$MeV is obtained.
We do  not want to stress here any qualitative comparison but rather
like to point out that our approach also leads to cut--off values
well below the ones obtained in OBE approaches.

Introducing a flat metric requires a canonical form
for the kinetic part of the lagrangian, which determines
the dynamics of the field fluctation. The kinetic term
of the scalar meson in Eq. \re{lagr}
${\cal L^{\mbox{kin}}_{\sigma}}= S_{0}^{2}e^{-2\sigma/d}  \pal_{\mu}\sigma
\pal^{\mu}\sigma/2  $
 can be easily rewriten in a  usual way:
${\cal L^{ \mbox{kin}  }_{\sigma}}= \pal_{\mu}\sigmat\pal^{\mu}\sigmat/2$
by the following redifinition of the basic sigma field:
$\sigmat(r)=S_0d[1-e^{-\sigma(r)/d}]$. Now the new field
$\sigmat$ may be identified with the real sigma field.
Clearly this redifinition does not change the nucleons static
properties given in Table~\ref{tab1}. Note also that, using
the above redinition in the last term of Eq. \re{lagr},
one may easily conclude that $m_{\sigmat}=m_{\sigma}$.
The appropiate sigma--nucleon form factor is given by
\be
G^{\sigma}(-{\vec q}^{\ 2})=-4\pi\ds\il_0^{\infty}
j_0( qr)\left[\sigmat^{\prime\prime}+\dsf{2\sigmat^{\prime}}{r}-
m_{\sigma}^2\sigmat\right]r^2dr,
\lab{ffs}
\ee
and may be used in OBE models. We have not introduced any metric
factors in the form factors of the heavier mesons since these
should play a lesser role than in the case of the pion.

 For small values of
the sqaured four--momentum transfer $t$
each form factor can be parametrized in monopole form:
 $G_{i}(t)=g_{i}(\Lambda_{i}^{2}-m_{i}^{2} )/(\Lambda_{i}^{2}-m_{i}^{2} )$
$(i=\pi, \rho, \omega, \sigma)$  . We present in Table~\ref{tab2}
 the range parameters
(cut--offs) and the coupling constants of the resulting
meson--nucleon dynamics. One can see that
there is no interference between $\sigma$--meson nucleon and e. g. the
pion--nucleon coupling constants. In other words,  the inclusion
of $\sigma$--meson does not significantly
affect meson--nucleon form factors that had been
given by the $\pirow$ model.
As it is seen from Table~\ref{tab2} the values for meson--nucleon
coupling constants are close to their emperical values (in some cases
obtained by OBE model fits).
This is one of the main advantages of the
inclusion of a scalar--isoscalar meson as done in the present approach.

The corollary of the present model is that
it gives  significant information
on the $\sigma$--nucleon interaction. As it is seen from the
Table~\ref{tab2},
the value for $ g_{\sigma NN}$ and  the cut--off parameter
of sigma --nucleon vertex $\Lambda_{\sigma}$
 are  smaller than their OBE prediction
 $\Lambda_{\sigma}^{OBE}\approx 1300\div 2000$ MeV.
    This contrast is
evidently seen from the  Fig.~2., where
$G^{\sigma}(t)/G^{\sigma}(0)$ for two cases : $\msig=720 $MeV
and  $\msig=550 $MeV is presented with the solid and dashed
 lines respectively. The band enclosed by the dotted lines refers to the OBE
monople form factor  with $\Lambda_{\sigma}^{OBE}=1300\ldots 2000\,$MeV.
  One can  conclude that the present model
gives a softer $\sigma NN$ form factor than it obtained by
OBE.  As it had been noticed before,  the  $t$--plane
for each form factor has a cut along the positive real axis extending from
$t=t_0$
to $\infty$.  The cut for the $\sigma$--nucleon vertex
function starts at $t_0=4\mpi^{2} $ reflecting the kinematical
threshold for the $\sigma\rightarrow\pi\pi$ channel.
More precisely this result follows from the asymptotic behavior
of the meson profiles: For $r\rightarrow\infty$, we have
$\Theta(r)\sim\exp(-\mpi r)/\mpi r$,
 $ \sigma (r)\sim\Theta ^2 (r)$, which are derived from the
equations of motion.

Once the vertex function of the corresponding meson--nucleon interaction
has been found, its appropriate contribution to the $ NN$
interaction may be easily calculated by using well known techniques from
OBE.  The detailed formulas are given elsewhere~\ci{machrep, meisnn}.
In particular, the contribution of the
$\sigma$--meson exchange to the central potential is given by
\be
V^{c}_{\sigma}(r)=\ds \il_0^{\infty}\frac{k^2 dk}{2\pi^2}
\frac{G^{2}_{\sigma NN} (k^2)}{k^2+\msig^2} \, j_0(kr)~.
\ee
The central $NN$ potential in  the $T=0,  S=1 $ state (the deuteron state)
is presented in Fig.~3 in comparison with Paris potential.
Our prediction is in  good agreement with the
emperical one. Note that the desired attraction  in the central
 $V_{NN}$ has before been obtained in the $\pirow$ model by means of 
two--meson exchange~\ci{meisnn}.

In conclusion it should be noted that we do not intend (expect)
to describe (cover) all $NN$ phase shifts staying only in the
framework of the present model. Besides other mesons, which are usually
included in OBE picture, the full model should also take in
account e.g. $N\Delta\rho$ couplings. In addition
the $2\pi$ exchange and its strong mixing with
$\sigma$ meson exchange (see e.g. ref.\ci{birse}) should be considered.
Another reason possible which limits the accuracy 
of $NN$ phase shifts in the present model is
that the $\sigma$NN coupling is not sensitive the mass of
sigma (see Table~\ref{tab2}) as it is in the OBE phenomenology. In fact, even
when the $2\pi$ exchange is disregarded , the pure OBE
model has to consider two types of sigmas
with nearly the same masses but with quite different coupling
constants. So, we refrain from performing direct calculations
of $NN$ phase shifts in the present model. Instead, we  point
out that    the meson--nucleon form factors found in the present
model could be useful in a wider context of calculations
of nucleon--nucleon observables (phase shifts, deuteron properties
etc) and may give  more information on meson--nucleon
and nucleon--nucleon dynamics.

To summarize, we have developed a topological chiral soliton model
with an explicit light scalar--isoscalar meson field, which
plays a central role in nuclear physics, based on the
chiral symmetry and broken scale invariance of QCD. We have shown that
for the single nucleon properties, the successfull description of the
electromagnetic observables of the $\pirow$ model is not modified and even
the value for the axial--vector coupling is somewhat improved. In the two--nucleon
sector, this extended $\pirows$ Lagrangian leads to the correct intermediate
range attraction in the central potential and a  soft $\sigma NN$
formfactor for   both values of sigma meson mass
 $\msig=550$ MeV and   $\msig=720$ MeV.

\bigskip
\bigskip
{\bf Acknowledgements:}
We would like to thank R.~Machleidt for providing us with computer codes
and M.~Musakhanov for useful discussions.
This work was supported in part by Deutscher Akademischer Austauschdienst
(DAAD).

\pagebreak

\renewcommand{\arraystretch}{1.1}
\newpage

\begin{table}[hbt]
\caption{
 Baryon properties in the $\pirow$ and $\piros$ models
}
\vspace {1cm}
\begin{center}
\begin{tabular}{|l|c|c|c|c|}\hline
           &       $\pirow$  &$\piros$     &$\piros$ & Exp. \\
\hline
$\msig$ [MeV] & $-$              &550          &720& \\
$C^{1/4}_{g}$ [MeV]  & $-$              & 258  & 295  &
     300$\div$400  \\
$B^{1/4}$  [MeV]  & $-$  & 119 & 121  & $-$       \\
$M_N$ [MeV] & 1560     & 1492 & 1511 &        939\\
$\Lambda$ [fm] & 0.88            & 0.88 & 0.88 &    $-$ \\
$M_\Delta-M_N$ [MeV]&    344  & 350  & 350  &        293 \\
$r_H=\langle r_B^2\rangle^{1/2}$ [fm]& 0.5  & 0.5  & 0.5  & $\sim$0.5  \\
$\langle r_E^2\rangle^{1/2}_p$ [fm] & 0.92 & 0.94 & 0.94 &  0.86$\pm$ 0.01\\
$\langle r_E^2\rangle_n$ [fm$^2$] & -0.20& -0.16&-0.16 & -0.119$\pm$0.004\\
$\langle r_M^2\rangle^{1/2}_p$[fm] & 0.84 & 0.85 & 0.85 & 0.86$\pm$ 0.06 \\
$\langle r_M^2\rangle^{1/2}_n$ [fm] & 0.85 & 0.85 & 0.85 & 0.88$\pm$ 0.07 \\
$\mu_p$ [n.m.]        & 3.34 & 3.33 & 3.33 &        2.79\\
$\mu_n$ [n.m.] & -2.58  & -2.53& -2.53&    -1.91 \\
$|\mu_p/\mu_n|$ & 1.29         & 1.30 & 1.30 &        1.46  \\
$g_A$ &0.88                  & 0.95 & 0.95 &  1.26$\pm$0.006\\
$\langle r_A^2\rangle^{1/2}_p$ [fm] & 0.63 & 0.66 & 0.66 &  0.65$\pm$ 0.07\\
\hline
\end{tabular}
\end{center}
\label{tab1}
\end{table}
\newpage
\begin{table} [hbt]
\caption{
Meson--nucleon coupling constants and cut--off parameters of
meson--nucleon form factors.
The $\Lambda_i(i=\pi, \rho, \omega\sigma)$ are cutoff parameters
in equivalent monopole fits $1/(1-t/\Lambda_i^2)$ to the
normalized  form factors
$G_{iNN}(t)/G_{iNN}(0)$
around $t=0$. The empirical
values are from OBE potential fit \ci{machrep}
}
\vspace{1cm}
\begin{center}
\begin{tabular}{|l|c|c|c|c|}\hline
           &         $\pirow$      &$\piros$(550)  &$\piros$(720)& OBE/Emp.\\
\hline
$G_{\pi NN}(0)$ &
14.74& 13.97 & 14.17 & 13.53\\
$G_{\sigma NN}(0)$ &
$-$ & 6.2   & 6.19  & 9.1 (12.41) \\
$F_1^\rho(0)$ &
2.55 & 2.76 & 2.68 & 2.24\\
$F_2^\rho(0)$ &
14.33 & 15.01 & 14.67 & 13.7  \\
$F_2^\rho(0)/F_1^\rho(0)   $ &
5.6      & 5.43  &5.47   & 6.1  \\
$F_1^\omega(0)$ &
8.78 & 10.73& 10.15& 11.7 \\
$F_2^\omega(0)$ &
$-2.15$ & $-2.78$ & $-2.65$ & 0 \\
$F_2^\omega(0)/F_1^\omega(0)   $ &
-0.24     &-0.25  &-0.26   & 0  \\
$\Lambda_\pi$ (GeV) &
1.2  & 1.1  & 1.1  & 1.3$\div$2.0 \\
$\Lambda_\sigma$ (GeV) &
$-$ & 0.59 & 0.60 & 1.3$\div$2.0    \\
$\Lambda^{\rho}_{1}$ (GeV) &
0.62   & 0.63 & 0.63 & 1.3 \\
$\Lambda^{\rho}_{2}$ (GeV) &
0.92   & 0.92 & 0.92 & 1.3 \\
$\Lambda^{\omega}_{1}$ (GeV) &
0.95     & 0.89 & 0.91 & 1.5 \\
$\Lambda^{\omega}_{2}$ (GeV) &
1.12   & 0.86 & 0.89 & - \\
\hline
\end{tabular}
\end{center}
\label{tab2}
\end{table}

\newpage



\begin{figure}[hbt]
   \vspace{0.5cm}
   \epsfysize=18cm
   \centerline{\epsffile{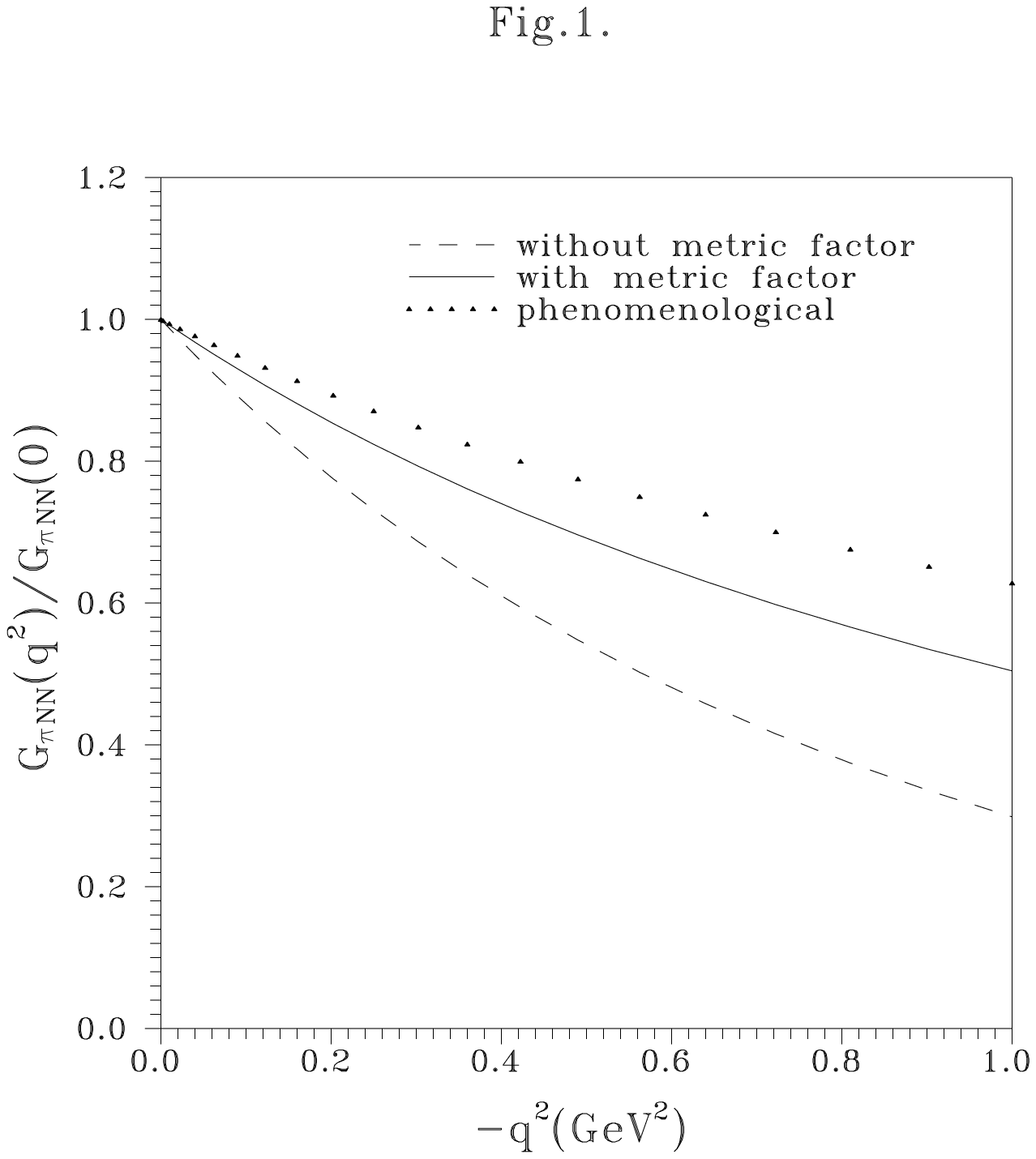}}
   \vspace{-3.0cm}
   \centerline{\parbox{15cm}{\caption{\label{fig1}
The normalized $\pi NN$  form factor in the
$\pi\rho\omega\sigma$ model ($\msig=720$ MeV).
The solid line represents the form factor when the metric factor is included
(Eq. \re{ffmet}), while the
dashed line gives the result with no metric factor as in Eq. \re{ff}. The
dotted line is a monopole form factor with
$\Lambda_{\pi}=1300\,$MeV.
  }}}
\end{figure}

\begin{figure}[htb]
   \vspace{0.5cm}
   \epsfysize=19cm
   \centerline{\epsffile{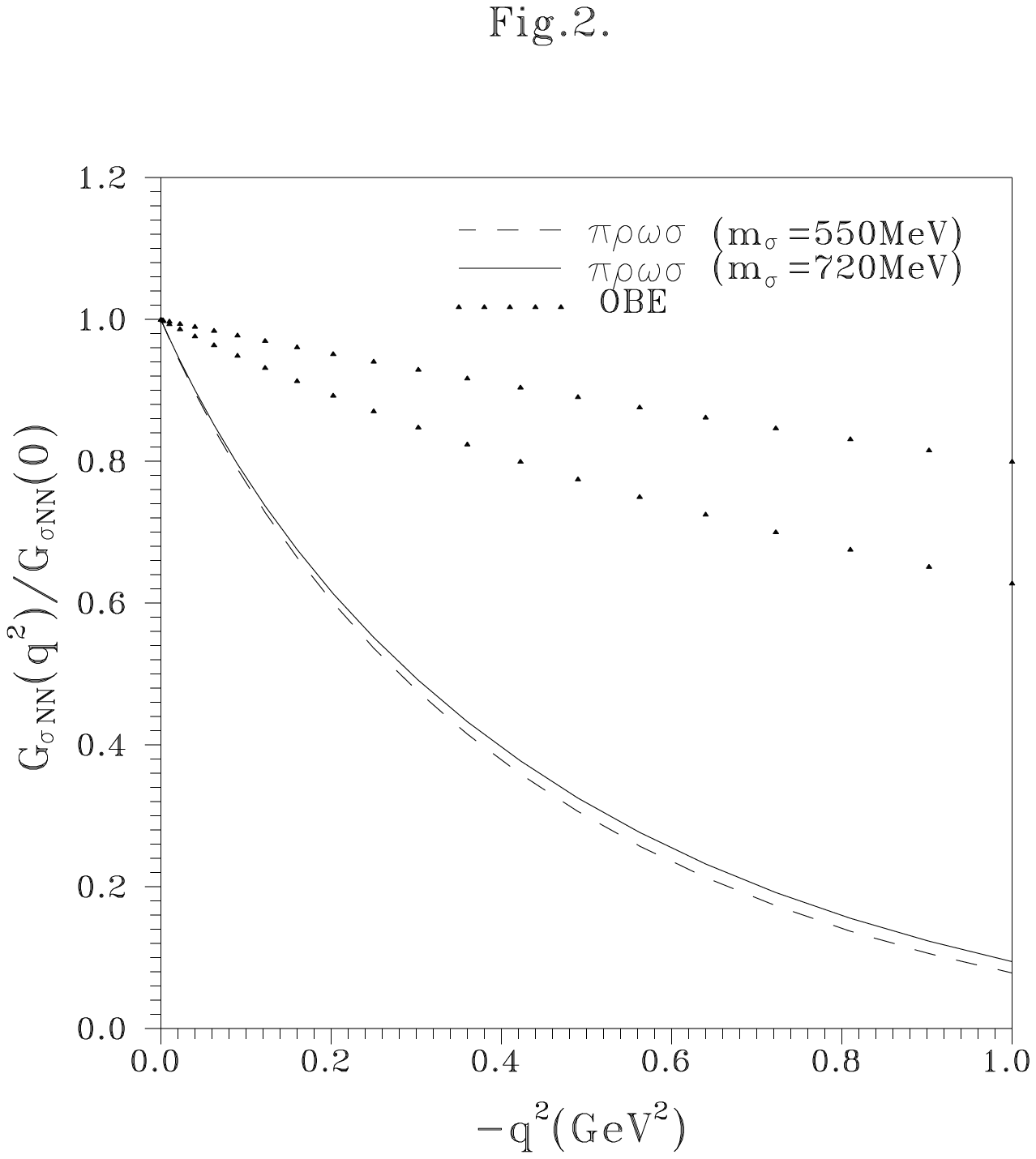}}
   \vspace{-3.0cm}
   \centerline{\parbox{15cm}{\caption{\label{fig2}
The sigma--nucleon form factor
 $G_{\sigma NN}(\vec q^{\,2})/G_{\sigma NN}(0)$.
The dashed and solid lines are for $m_{\sigma}=550$~MeV and
$m_{\sigma}=720$~MeV, respectively. Typical OBE monopole fits with
 $\Lambda = 1.3 \ldots 2$ GeV are shown by 
the band enclosed by the dotted lines.}}}
\end{figure}

\begin{figure}[htb]
   \vspace{0.5cm}
   \epsfysize=19cm
   \centerline{\epsffile{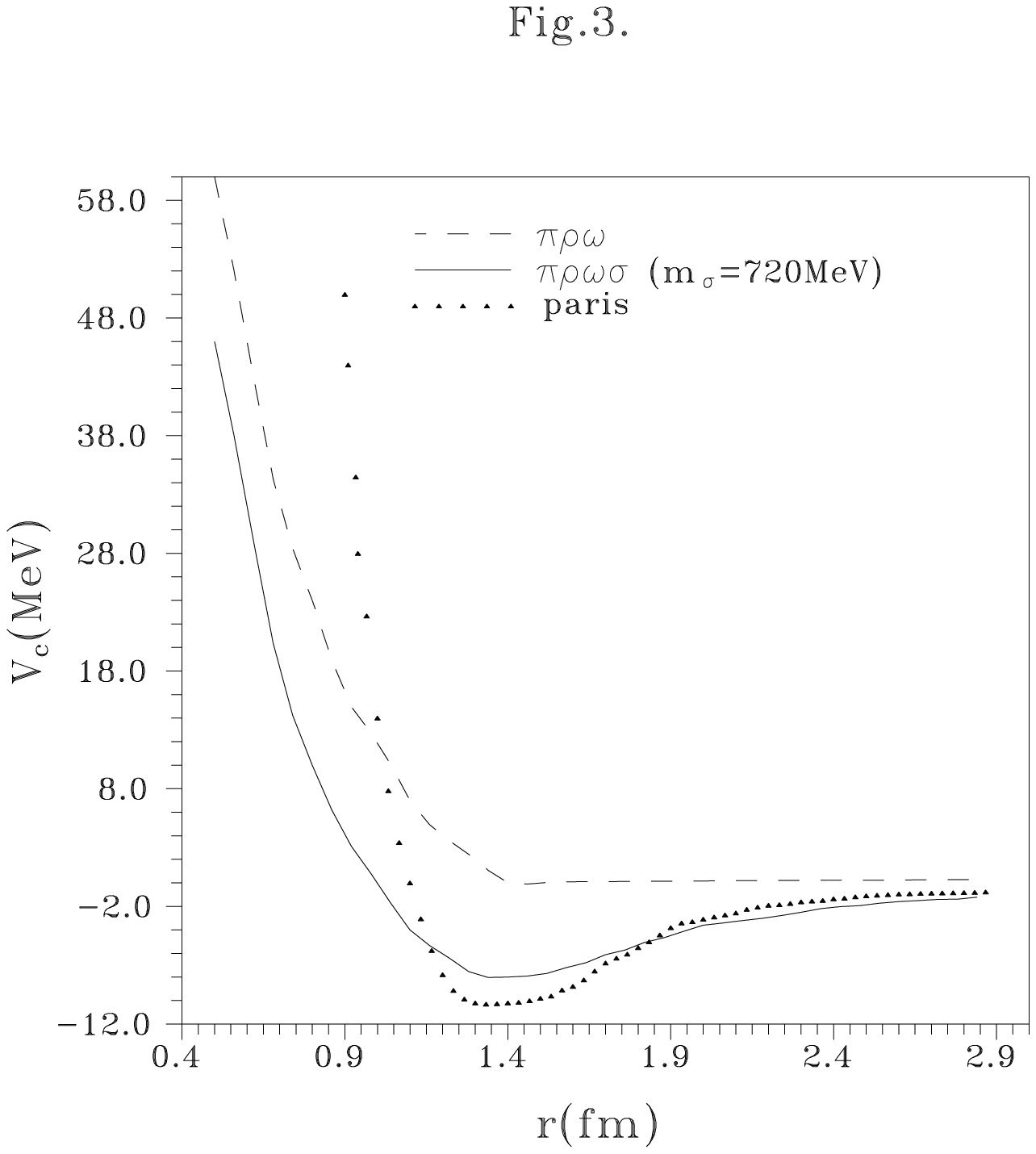}}
   \vspace{-3.0cm}
   \centerline{\parbox{15cm}{\caption{\label{fig3}
The central potential in the $S=1$, $T=0$ state for
$\pi\rho\omega$ and $\pi\rho\omega\sigma$ models (dashed and solid lines,
respectively). No contribution  from two--meson exchange has been taken
into account. The dotted line corresponds to the Paris
potential~{\protect \ci{paris}}.
  }}}
\end{figure}

\end{document}